\documentclass[a4paper,11pt]{article}
\usepackage{jheppub} 
\usepackage{mathtools}
\usepackage{slashed}
\usepackage{booktabs}
\usepackage{tabularx}
\usepackage{array}
\usepackage{amscd}
\usepackage{float}
\usepackage{xcolor}
\usepackage{tikz-feynman}
\usetikzlibrary{positioning}
\tikzfeynmanset{compat=1.1.0}

\newcommand{\SU}{\mathrm{SU}}

\newcommand{\U}{\mathrm{U}}

\newcommand{\AdS}{\mathrm{AdS}}
\newcommand{\CFT}{\mathrm{CFT}}

\newcommand{\Pl}{\mathrm{Pl}}
\newcommand{\Tr}{\operatorname{Tr}}

\newcommand{\dd}{\mathrm{d}}

\newcommand{\order}{\mathcal O}

\title{Making the Great Desert Bloom}

\author{Jacques Distler and}
\author{Andreas Karch}
\affiliation{Theory Group, Weinberg Institute, Department of Physics,
University of Texas \\  2515 Speedway, Austin, TX 78712, USA.}

\emailAdd{distler@golem.ph.utexas.edu,karcha@utexas.edu}

\abstract{The absence of new physics beyond neutrino masses between the weak scale and the Planck scale is referred to as the great desert. It is often seen as a nightmare scenario for particle physics. Here we argue, by contrast,  that the great desert assumption in fact has dramatic implications: it predicts that the Standard Model has a semi-classical AdS$_4$ vacuum with CFT$_3$ dual and, under reasonable assumptions, must be accompanied by at least one extra dimension significantly larger than Planck length. We also present one plausible candidate CFT$_3$ to be ``the" dual to the Standard Model.}

\begin{document}
\maketitle
\flushbottom

\section{Introduction}
The great desert hypothesis says that, after the Standard Model is supplemented by whatever is needed for neutrino masses and by gravity, no new non-gravitational threshold appears between the electroweak scale and the Planck scale.  This is often regarded as a worst-case outcome for particle physics, because it gives no obvious new particles, no low-energy symmetry principle, and no collider-accessible guide to ultraviolet physics.

In this work we want to take the great desert seriously and emphasize, that in contrast to naive expectations, it in fact leads to very interesting physics. With measured Higgs, top, and strong-coupling inputs, the Higgs quartic runs negative at high field placing the electroweak vacuum in the metastable regime~\cite{Isidori:2001bm,Degrassi:2012ry,Buttazzo:2013uya} and leading to a run-away behavior at energies well below the Planck scale. Once the theory is coupled to gravity, Planck-suppressed operators need to be included in the effective theory \cite{Donoghue:1994dn,Isidori:2007vm}.  If the leading sextic correction to the Higgs potential has the sign that gives a controlled high-field vacuum, the vacuum energy of that minimum is negative. Thus the great desert naturally leads not merely to our electroweak vacuum, but also to a high-VEV Standard-Model AdS$_4$ vacuum with curvature radius large-ish in Planck units.

This observation has dramatic consequences. Holography demands that the large Higgs AdS$_4$ vacuum has a CFT$_3$ dual. Note that once one correctly model-builds this 3d CFT one not only gets a full UV completion of the Standard Model coupled to gravity, one also predicts the masses and interactions of all particles in the Standard Model from first principles. Some of the properties of the putative dual are directly predicted by the known physics of the Standard Model run up to the large Higgs VEV. $SU(2)\times U(1)$ is broken  (even more strongly than it is in our vacuum). Because the AdS radius is tiny compared to the QCD confinement scale, bulk photons and gluons are massless, unconfined gauge fields, dual to conserved currents in the CFT. The light bulk fermions of the first and second generation are dual to  fermionic operators in the CFT with \(\Delta\sim3/2\).  The stress-tensor normalization, $C_T$,  is large since the AdS radius is large in Planck units. 

While we have no reason to expect any light bulk scalars, the mere existence of a CFT$_3$ dual strongly suggests that the great desert in fact is accompanied by large-ish extra dimensions of size comparable to the AdS$_4$ curvature radius.

Bootstrap bounds tell us that low dimension fermions cannot exist without low dimension scalars \cite{IliesiuActual}. However, in a large-\(N\) CFT with a weakly coupled AdS dual, the bootstrap bounds can easily be satisfied  by fermion-bilinear double-trace operators. Bootstrap bounds alone do not constrain an otherwise consistent bulk effective theory \cite{Heemskerk:2009pn}. 
Nevertheless, a generic large-\(N\) holographic CFT is expected to contain infinitely many single-trace primaries, reflecting the UV completion of the bulk, with, barring any strong coupling magic, dimensions comparable to those of the low-lying fermions. This is certainly the case in our candidate CFT$_3$ dual for the Standard Model. Such a tower of single trace primaries strongly suggests the existence of at least one large extra dimension in the bulk. This is surely what one expects in a generic large $N$ CFT. Whether lack of scale separation between an AdS vacuum and the size of the extra dimension is, in fact, a principle obeyed by {\it any} consistent theory of quantum gravity (and hence its dual CFT) is an area of active debate, for a review see for example \cite{Coudarchet:2023mfs}. 

Note that a different kind of CFT dual to the standard model has previously been proposed in \cite{Arkani-Hamed:2007ryu}. There it was argued that the standard model should also allow for an AdS$_3$ $\times$ $S^1$ vacuum with the small circle stabilized by Casimir energies. This did not rely on the metastability of our vacuum  and the resulting dual 2d CFT is very distinct from the dual we are exploring here. Let us also note that our results are in clear clash with one of the swampland conjectures postulating the absence of non-supersymmetric AdS vacua \cite{Ooguri:2016pdq}. In the absence of a full string theory embedding it is not obvious exactly how our construction avoids these arguments, but given the conjectural nature of the bound it should not be taken as a show-stopper. Conversely, if the bound of \cite{Ooguri:2016pdq} is in fact a property of any consistent theory of quantum gravity, our results show that it basically rules out the great desert scenario.

This paper is organized as follows: in section \ref{sec:highVEV} we review the known physics of the Higgs potential under the great desert assumption and exhibit the properties of the resulting AdS$_4$ vacuum. In section \ref{sec:towards} we spell out our assumptions. The great desert together with the assumption of a stable vacuum immediately predict the existence of a CFT dual. One further mild assumption then motivates at least one large extra dimension, and a fourth assumption about the neutrino mass mechanism pins down the global symmetry of the CFT. In section \ref{sec:cftcandidate} we present a CFT$_3$ candidate and analyze its light operators, its larger single-trace spectrum, and its current contact terms.  We conclude in section \ref{sec:discussion} with open questions and suggestions for a microscopic completion.

\section{The high-VEV Standard Model \texorpdfstring{AdS$_4$}{AdS₄} vacuum  }
\label{sec:highVEV}

\subsection{AdS$_4$ from higher dimension operators}

At large Higgs field VEV $  \langle H \rangle  = (0,|H|)$ the electroweak quadratic term is negligible and the RG-improved potential is
\begin{equation}
  V_{\rm SM}(|H|)=\lambda_{\rm eff}(|H|) |H|^4 .
\end{equation}
For measured Standard-Model inputs, \(\lambda_{\rm eff}\) crosses zero at a high scale around $10^{10}$ GeV, well below the Planck scale~\cite{Degrassi:2012ry,Buttazzo:2013uya}. It reaches a minimal value 
\begin{equation}
\lambda_* \approx - 0.013 
\end{equation}
at an energy between $10^{17}$ and $10^{18}$ GeV with very little running near that energy range, 
The pure Standard Model therefore has no stable vacuum in the controlled region; it runs toward negative energy.

As the Higgs VEV approaches Planckian values higher order terms in the effective field theory become important. These higher order corrections can stabilize the Standard Model vacuum. The leading order correction to the Higgs potential is given by
\begin{equation}
  V(|H|)=\lambda_{\rm eff}(|H|)|H|^4+ 8 c_6\frac{|H|^6}{M_{\Pl}^2}+\cdots .
  \label{eq:sexticpotential}
\end{equation}
where the running $\lambda_{\rm eff}$ is a function of $|H|$. If $c_6$ is positive the higher order correction results in a stable minimum at scales near but still clearly below the Planck scale.
Since the minimum of $V(|H|)$ occurs near the minimum of $\lambda_{\rm eff}$, and since $\lambda$ is a very  slowly running function around that scale, to a very good approximation we can replace $\lambda_{\rm eff}$ with its minimal value $\lambda_*$ near the minimum $|H|=v_2/\sqrt{2}$ of $V(|H|)$. With this we find that at the minimum
\begin{equation}
  v_2=M_{\Pl}\sqrt{\frac{|\lambda_*|}{6c_6}}
  \label{eq:v2}
\end{equation}
for \(c_6>0\).  The corresponding vacuum energy is
\begin{equation}
  V_{\rm min}=-\frac{|\lambda_*|}{12}v_2^4<0.
  \label{eq:vmin}
\end{equation}
Thus the controlled high-field minimum is an \(\AdS_4\) vacuum.

The AdS radius is
\begin{equation}
  L=\sqrt{\frac{3M_{\Pl}^2}{|V_{\rm min}|}}
   =\frac{6M_{\Pl}}{|\lambda_*|^{1/2}v_2^2}
   =\frac{36c_6}{|\lambda_*|^{3/2}M_{\Pl}}.
  \label{eq:Lads}
\end{equation}
For \(|\lambda_*|\simeq0.013\), using the reduced Planck mass
\(M_{\Pl}=2.4\times10^{18}\,\mathrm{GeV}\),
\begin{equation}
  LM_{\Pl}\simeq2.4\times10^4 c_6
\end{equation}
The semiclassical regime is therefore compatible with a wide range of \(c_6\). A good measure for the effective number of degrees of freedom is
\begin{equation}
  C_T\sim (LM_{\Pl})^2\simeq5.9\times10^8c_6^2.
  \label{eq:CT}
\end{equation}
  where $C_T$ is the normalization of the stress tensor 2-pt function in the putative dual CFT and should be thought of as a stand in for $N^2$, where $N$ is a large integer counting degrees of freedom in the dual CFT. For \(c_6\sim1\) the putative dual CFT is very large-\(N\); for \(c_6\sim10^{-2}\) it is still plausibly semiclassical. 

\subsection{Broken and unbroken gauge symmetries}

In ``our" vacuum with Higgs VEV $v_1$ we know that $SU(2) \times U(1)_Y$ is broken to $U(1)_{EM}$ by the Higgs VEV whereas the $SU(3)$ of QCD confines. In the large-Higgs VEV vacuum with $v_2 \gg v_1$ $SU(2)$ is broken much more strongly and the only surviving gauge symmetry from the electroweak sector is $U(1)_{EM}$. In contrast, QCD is actually de-confined in the large Higgs VEV vacuum. AdS$_4$ acts as a box of size $L$, providing an effective infrared cutoff. If the box is very small compared to the QCD scale $\Lambda^{-1}$, which is clearly the case here, asymptotic freedom of QCD implies that QCD is very weakly coupled and in particular deconfined. So in the large-Higgs VEV vacuum the usual Standard Model gauge symmetries lead to an $SU(3) \times U(1)_{EM}$ gauge symmetry with the corresponding eight massless gluons and one massless photon.

To complete the discussion of symmetries, we have to commit to a mechanism for the neutrino masses. Strictly speaking, the Standard Model as written requires neutrinos to be massless and the non-vanishing masses of the neutrinos seen in experiments already force us to include some beyond-the-Standard Model physics even in our great desert scenario. We read the great desert as stating that there is no new physics below the Planck scale {\it other} than that demanded by neutrino masses. But since, as of now, we do not know the detailed mechanism for neutrino masses we have to make a choice of how we believe neutrinos get their mass.

The simplest case to discuss is a type I seesaw completion with gauged\footnote{That \(\U(1)_{B-L}\) is gauged accords  with the general dogma that there should be no ungauged global symmetries in quantum gravity.} \(\U(1)_{B-L}\), three right-handed neutrinos, and a complex scalar \(\phi\) of \(B-L\) charge \(+2\).  In our vacuum $\phi$ has a large VEV $v_{\phi}/\sqrt{2}$ giving rise to a Majorana mass term $M_R$ $\propto$ $v_{\phi}$. For each family this gives rise to the light observed neutrino with mass $\propto$ $v_1^2/v_{\phi}$ and a heavy partner with mass $\propto v_{\phi}$.

Including $\phi$, our scalar potential takes the form
\begin{equation}\label{eq:portal}
\begin{split}
V(H,\phi)={}&
\lambda_1\left(|H|^2-\tfrac{1}{2}v_1^2\right)^2
+\lambda_2
\left(|\phi|^2-\tfrac{1}{2}v_\phi^2\right)^2
\\
&+\lambda_3
\left(|H|^2-\tfrac{1}{2}v_1^2\right)
\left(|\phi|^2-\tfrac{1}{2}v_\phi^2\right)
+\frac{8 \hbar c_6}{M_{\rm Pl}^2}
\left(|H|^2-\tfrac{1}{2}v_1^2\right)^3 .
\end{split}
\end{equation}
where we've put an explicit $\hbar$ in front of the $c_6$ term to remind us that, in the EFT power-counting, this quantum gravity correction counts as a 1-loop correction, contributing at the same order as the 1-loop effective potential which, in the great desert scenario, drives $|H|$ to large VEV.

Here we have a binary choice: what is the sign of $\lambda_3$? Let us first assume it is positive. In that case, for fixed $|H|$, the minimum of the potential is located at
\begin{equation}
0=\phi \left[|\phi|^2 -\tfrac{1}{2}v_\phi^2 +\frac{\lambda_3}{ 2\lambda_2} \left(|H|^2 -\tfrac{1}{2}v_1^2\right)\right]
\end{equation}
 For large Higgs VEV, this drives $\phi\to 0$. Hence, in the true vacuum, \(\U(1)_{B-L}\) is unbroken and the neutrinos are Dirac particles with mass $\propto v_2$. Of course, that conclusion holds only if
 \begin{equation}
\frac{\lambda_3}{\lambda_2} \gtrsim 2 \frac{v_\phi^2}{v_2^2}
\end{equation}
which is the condition that
\begin{equation}
m_{\phi}^2 \approx \frac{\lambda_3}{2}   v_2^2 - \lambda_2 v_{\phi}^2 
\end{equation}
is positive in the true vacuum.
More usefully, we can write this as 
\begin{equation}
\label{eq:phimass}
(m_{\phi}L)^2 \approx \frac{108c_6}{|\lambda_*|^2}\left[\lambda_3 - 12 \frac{c_6 \lambda_2}{|\lambda_*|}\frac{v_\phi^2}{M_{\Pl}^2}\right]
\end{equation}
Since $v_\phi/M_{\Pl}\sim O(10^{-4})$, the factor in square brackets is $\lambda_3$, unless the latter is fine-tuned to be unnaturally-small.  So, overall, we expect
\begin{equation}
m_\phi L \approx 10 \frac{\sqrt{c_6 \lambda_3}}{|\lambda_*|}
\end{equation}
For $c_6\sim\lambda_3\sim O(1)$, this is $O(10^3)$, but could be as small as $O(10)$ for $c_6\sim\lambda_3\sim O(10^{-2})$. So it is not implausible (though it requires a mild fine-tuning)  that $\phi$ is relatively light, in AdS units,  in the true vacuum.


That $U(1)_{B-L}$ is unbroken in the true vacuum has two consequences.  First, the \(U(1)_{B-L}\) gauge boson is massless in the AdS vacuum and gives an additional conserved current in the CFT.  The target global symmetry of the CFT is therefore
\begin{equation}
  \SU(3)_c\times\U(1)_{\rm EM}\times\U(1)_{B-L}.
  \label{eq:targetsym}
\end{equation}
(We will be more careful about the global form of the bulk gauge group, and hence of the flavour symmetry of the CFT, below.)

Second, the same ultraviolet neutrino Yukawa matrix has a different physical
interpretation in the two vacua.  In our vacuum, $\langle\phi\rangle=v_\phi/\sqrt{2}$
generates a Majorana mass matrix $M_R$, and integrating out the right-handed
neutrinos gives
\begin{equation}
  m_\nu^{(1)}
  =
  -\frac{v_1^2}{2}\,Y_\nu^T M_R^{-1}Y_\nu .
\end{equation}
For example, $M_R\sim10^{14}\,\mathrm{GeV}$ together with a nonzero singular
value $y_\nu\sim0.4$ gives a light-neutrino mass of order
$5\times10^{-2}\,\mathrm{eV}$.
In the high-VEV branch considered here, by contrast,
$\langle\phi\rangle=0$, so $M_R=0$ and the right-handed neutrinos remain in
the spectrum.  Their masses are the singular values of the Dirac mass matrix
\begin{equation}
  m_D(v_2)=\frac{v_2}{\sqrt2}\,Y_\nu(v_2).
\end{equation}
If $Y_\nu$ has full rank, all three neutrinos are heavy Dirac fermions; a singular value $y_\nu\simeq0.4$ of $Y_{\nu}$ gives $m_\nu\simeq0.3v_2$.  The CFT
candidate studied below is instead intended to describe the rank-two
possibility
\begin{equation}
  \operatorname{rank}Y_\nu(v_2)=2,
\end{equation}
for which one neutrino remains massless (or nearly massless) in the high scale vacuum while the
other two are heavy.  

The rank-two structure is stable under the standard one-loop
renormalization-group evolution.  In the effective theory below the seesaw
thresholds, the first unavoidable Standard-Model contribution capable of
lifting the zero mass eigenvalue appears at two loops
\cite{Antusch:2001ck,Antusch:2005gp,Ibarra:2024WeinbergRGE}.
For a matching scale of order \(10^{14}\,\mathrm{GeV}\), the resulting
radiative mass is only
\begin{equation}
  m_1^{\rm rad}\sim 10^{-13}\,\mathrm{eV}
\end{equation}
in the Standard Model
\cite{Davidson:2006tg,Xing:2020ezi}.
That is a rank-2 mass matrix at high scale gives one abnormally light neutrino in our standard model vacuum. In normal
ordering, the observed mass-squared differences then give
\begin{equation}
  m_2\simeq 8.6\times10^{-3}\,\mathrm{eV},
  \qquad
  m_3\simeq 5.0\times10^{-2}\,\mathrm{eV},
\end{equation}
while \(m_1^{\rm rad}\) is about eleven orders of magnitude smaller.
Since oscillation experiments determine mass-squared differences rather
than the absolute lightest mass, this spectrum is fully consistent with
current data \cite{PDGNeutrino2025}.  Additional threshold effects,
irrelevant operators, or ultraviolet sources of rank breaking could lift
\(m_1\) above this radiative floor.

In this work we mostly focus on this $\lambda_3>0$ branch, in which $U(1)_{B-L}$ is
unbroken in the high-VEV vacuum.  Its bulk gauge group, and hence the global
symmetry of the putative dual CFT, is\footnote{More precisely, the faithfully acting global form relevant below is \begin{equation} G_{v_2} = \frac{SU(3)_c\times U(1)_{1}}{\mathbb Z_3} \times U(1)_{2}= U(3)\times U(1)_{2}. \label{eq:highVEVglobalgroup} \end{equation} Here, the generator of $U(1)_1$, $Q_1=3(B-L)$. The generator of $U(1)_2$ is $Q_2=B-L+Q_{EM}$. The Lie algebra is of course $\mathfrak{su}(3)_c\oplus \mathfrak{u}(1)_{\rm EM}\oplus \mathfrak{u}(1)_{B-L}$; the $\mathbb{Z}_3$ quotient determines the global charge lattice, so that operators in the fundamental representation of $SU(3)$ carry $B-L\in \mathbb{Z}+1/3$ and all operators have $Q_2\in \mathbb{Z}$. This will be important for the contact-term analysis in Sec.~\ref{sec:newcontacts}.}
\begin{equation}
  G_{v_2}
  =
  SU(3)_c\times U(1)_{\rm EM}\times U(1)_{B-L},
\end{equation}
with ten massless gauge bosons; in the rank-two neutrino option the spectrum
also contains one massless neutrino. 

For the opposite sign,
$\lambda_3<0$, $\langle\phi\rangle$ tracks $\langle H\rangle$, so the Dirac
and Majorana masses scale together and the seesaw remains operative.  That
branch has only $SU(3)_c\times U(1)_{\rm EM}$, with nine massless gauge
bosons, and three potentially light neutrinos.  It is equally possible in
principle, but we do not pursue it below.

\subsection{Particle Masses and operator dimensions}\label{particleMasses}

We can use our knowledge of the Standard Model masses and couplings in our vacuum to deduce the particle spectrum in the large VEV AdS$_4$ vacuum as a function of the free parameter $c_6$ by running the couplings up to a high scale. In AdS, the meaningful quantity to consider is $mL$, the mass in units of the AdS curvature radius. Since couplings run slowly the main difference between different values of $c_6$ is simply the rescaling of $L$ according to \eqref{eq:Lads}. In the putative dual CFT $mL$ is directly related to the dimension of the dual operator.

A bulk particle of mass \(m=a v_2\) has
\begin{equation}
  mL=\frac{6\sqrt{6c_6}}{|\lambda_*|}a\simeq1131\,a\sqrt{c_6}.
  \label{eq:mL}
\end{equation}
The coefficient \(a\) is not simply the observed low-energy mass divided by the electroweak VEV.  It is obtained by evaluating the appropriate running Standard-Model coupling at the high-field scale \(\mu\simeq v_2\):
\begin{equation}
  a_W=\frac{g_2(v_2)}{2},
  \qquad
  a_Z=\frac{\sqrt{g_2^2(v_2)+g'^2(v_2)}}{2},
  \qquad
  a_f=\frac{y_f(v_2)}{\sqrt2}.
  \label{eq:adefinitions}
\end{equation}
We use the high-scale Standard-Model running of Refs.~\cite{Degrassi:2012ry,Buttazzo:2013uya}.  For the light-fermion Yukawas we use the updated two-loop running table of Ref.~\cite{Antusch:2025fpm}, which gives \(\overline{\rm MS}\) Yukawa and gauge couplings through \(10^{16}\,\mathrm{GeV}\); the final one or two decades up to \(v_2\) are obtained by one-loop SM running.  This last step changes the entries only mildly because the couplings run slowly in this range.  We use GUT normalization for \(g_1\) in the RGE input and convert to the hypercharge coupling by \(g'=\sqrt{3/5}\,g_1\) in Eq.~\eqref{eq:adefinitions}.

For the Higgs itself, more precisely,
\begin{equation}
  m_h^2=\left(2|\lambda_*|+\frac14\beta_\lambda'\right)v_2^2,
\end{equation}
where \(\beta_\lambda'=\dd\beta_\lambda/\dd\ln\mu\).  In the same benchmark approximation,
\begin{equation}
  m_h\simeq\sqrt{2|\lambda_*|}\,v_2\simeq0.161v_2,
  \qquad
  m_hL\simeq182\sqrt{c_6}.
  \label{eq:mHL}
\end{equation}
Thus for generic \(c_6=\order(1)\), the Higgs operator is heavy in AdS units.

The entries in Table~\ref{tab:mLdimensions} are obtained by combining Eq.~\eqref{eq:mL} with the standard \(\AdS_4/\CFT_3\) mass--dimension map.  For a scalar, spinor, and massive vector respectively,
\begin{align}
  \Delta_0&=\frac32+\sqrt{\frac94+m^2L^2},\\
  \Delta_{1/2}&=\frac32+|m|L,\\
  \Delta_1&=\frac32+\sqrt{\frac14+m^2L^2}.
\end{align}
A massless bulk gauge field is dual to a conserved current with \(\Delta=2\).  The two displayed columns use \(\mu=v_2\simeq1.12\times10^{17}\,\mathrm{GeV}\) for \(c_6=1\) and \(\mu=v_2\simeq1.12\times10^{18}\,\mathrm{GeV}\) for \(c_6=10^{-2}\).  

\begin{table}[H]
\centering
\scriptsize
\setlength{\tabcolsep}{3.2pt}
\renewcommand{\arraystretch}{1.08}
\caption{Dimensionless AdS masses and corresponding \(\CFT_3\) operator dimensions for two representative sextic coefficients, using \(|\lambda_*|=0.013\).  The coefficients \(a_i=m_i/v_2\) are evaluated from running Standard-Model couplings at the corresponding high-field scale. In addition, we display the mass of $\phi$, the scalar with $U(1)_{B-L}$ charge 2, for $\lambda_3=1$ and $\lambda_3=10^{-2}$.}
\label{tab:mLdimensions}
\vspace{0.35em}
\begin{tabular}{@{}lcccccc@{}}
\toprule
& \multicolumn{3}{c}{\(c_6=\lambda_3=1,\; v_2\simeq1.12\times10^{17}\,\mathrm{GeV}\)}
& \multicolumn{3}{c}{\(c_6=\lambda_3=10^{-2},\; v_2\simeq1.12\times10^{18}\,\mathrm{GeV}\)}\\
\cmidrule(lr){2-4}\cmidrule(l){5-7}
Particle & \(a_i\) & \(mL\) & \(\Delta\) & \(a_i\) & \(mL\) & \(\Delta\)\\
\midrule
\(W^\pm\) & 0.258 & 292 & \(\simeq293\) & 0.255 & 28.8 & 30.3\\
\(Z^0\) & 0.344 & 389 & \(\simeq390\) & 0.345 & 39.0 & 40.5\\
\(t\) & 0.304 & 344 & \(\simeq346\) & 0.295 & 33.4 & 34.9\\
\(h\) & 0.161 & 182 & \(\simeq184\) & 0.161 & 18.2 & 19.8\\
\(\tau\) & \(6.73\times10^{-3}\) & 7.61 & 9.11 & \(6.65\times10^{-3}\) & 0.752 & 2.25\\
\(b\) & \(4.12\times10^{-3}\) & 4.66 & 6.16 & \(3.97\times10^{-3}\) & 0.449 & 1.95\\
\(c\) & \(9.87\times10^{-4}\) & 1.12 & 2.62 & \(9.52\times10^{-4}\) & 0.108 & 1.61\\
\(\mu\) & \(3.96\times10^{-4}\) & 0.448 & 1.95 & \(3.92\times10^{-4}\) & 0.0443 & 1.54\\
\(s\) & \(8.81\times10^{-5}\) & 0.0996 & 1.60 & \(8.53\times10^{-5}\) & 0.00964 & 1.51\\
\(d\) & \(4.44\times10^{-6}\) & \(5.02\times10^{-3}\) & 1.505 & \(4.30\times10^{-6}\) & \(4.86\times10^{-4}\) & 1.500\\
\(u\) & \(1.95\times10^{-6}\) & \(2.21\times10^{-3}\) & 1.502 & \(1.89\times10^{-6}\) & \(2.13\times10^{-4}\) & 1.500\\
\(e\) & \(1.88\times10^{-6}\) & \(2.13\times10^{-3}\) & 1.502 & \(1.86\times10^{-6}\) & \(2.10\times10^{-4}\) & 1.500\\
\(\nu_{\rm heavy}\) & \(\simeq0.28\) & 317 & \(\simeq318\) & \(\simeq0.28\) & 31.7 & 33.2\\
\(\nu_0\) (rank 2) & 0 & 0 & 1.5 & 0 & 0 & 1.5\\
\(\gamma\) & 0 & 0 & 2 & 0 & 0 & 2\\
\(g\) (\(\times8\)) & 0 & 0 & 2 & 0 & 0 & 2\\
\(\phi\)& - & 799 &797&-&7.99& 6.23\\
\bottomrule
\end{tabular}
\end{table}

\section{Towards a CFT dual and the case for higher dimensions}
\label{sec:towards}

We laid out a clear cut case so far that under two basic assumptions
\begin{enumerate}
    \item Great desert: no new physics beyond neutrino masses up to near the Planck scale; the Standard Model plus gravity remains valid as an effective field theory
    \item This effective field theory has a stable vacuum
\end{enumerate}
the true vacuum of the Standard Model is a semi-classical AdS$_4$ and so it has a large $N$ CFT$_3$ dual. Let us pause a moment to emphasize how remarkable this is. If we can find the right CFT, not only will it give a UV completion of the Standard Model coupled to gravity. It will do so in a unique way, that is it will predict all Standard Model masses and couplings in terms of, at best, a finite set of numbers such as rank of the gauge group and Chern-Simons levels. The great desert is anything but boring! 

The CFT$_3$ is strongly constrained by symmetry. At minimum the global symmetry of this CFT has to include
\begin{equation}
G_{v_2}^0 = SU(3) \times U(1)_{EM} .
\end{equation}
This high degree of symmetry makes this CFT a very natural target for a bootstrap study. Unlike most 3d CFTs currently targeted by the bootstrap program, our CFT$_3$ is not parity invariant.

As we laid out in the previous section, including neutrino masses leads to essentially two options, depending on the sign of the coupling in \eqref{eq:portal}. While both scenarios can easily be studied, in this work we focus on one of the two scenarios and so make the additional assumptions
\begin{enumerate}
\setcounter{enumi}{2}
\item $U(1)_{B-L}$ is restored at large Higgs VEV
\end{enumerate}
This enlarges our target global symmetry to
\begin{equation}
G_{v_2} = SU(3) \times U(1)_{EM} \times U(1)_{B-L}.
\end{equation}

Last but not least, while not strictly speaking forced upon us by bootstrap bounds, it is very difficult to imagine a large $N$ field theory with the low dimension operator spectrum laid out in table \ref{tab:mLdimensions} without including an extra tower of light scalars. Here by light we mean masses with $mL$ an order 1 number, which is much lighter than Planck scale. These extra operators will definitely be present in our best-guess model presented in the next section, unless some unforeseen strong coupling magic pushes all of them up to very high dimension. The easiest way to accommodate these extra scalars is by the following assumption
\begin{enumerate}
\setcounter{enumi}{3}
\item The dual CFT has a tower of low dimension operators necessitating at least one large extra dimension of size comparable to $L$.
\end{enumerate}
Let us emphasize that while we in this work make all 4 assumptions in order to narrow down the physics enough to present a single model, assumptions 1 and 2 are already sufficient to lead to a CFT$_3$ dual for the Standard Model.

\section{A CFT candidate and its operator spectrum}
\label{sec:cftcandidate}

\subsection{Matter content, interactions, and the light fermions}
\label{sec:cftmatter}

The field theory we propose is a large-$N$ Chern--Simons matter theory whose connected global symmetry we engineer to be the desired 
\begin{equation}\label{eq:Fsym}
  SU(3)_c\times U(1)_{\rm EM}\times U(1)_{B-L}.
\end{equation}
We take the Chern-Simons gauge group to be
\begin{equation}
  G_{\rm CFT}=SO(2N_o)_{k_o}\times USp(2N_s)_{k_s}.
  \label{eq:OSpgauge}
\end{equation}
A natural choice is $N_o=N_s$ and the ABJ-like \cite{Aharony:2008gk,Hosomichi:2008jb} level assignment is $(k_o,k_s)=(2k,-k)$, but none of the kinematic statements below require this special locus.  Let
\begin{equation}
  {\cal R}=({\bf 2N_o},{\bf 2N_s}),\qquad D\coloneqq\dim{\cal R}=4N_oN_s,
\end{equation}
be the bifundamental representation.  The invariant tensors of the two factors combine to an antisymmetric invariant
\begin{equation}
  {\cal J}_{a\alpha,b\beta}=\delta_{ab}\Omega_{\alpha\beta},
  \qquad {\cal J}^T=-{\cal J},
  \label{eq:Jorthosymp}
\end{equation}
so ${\cal R}$ is pseudoreal.  We denote the antisymmetric bilinear pairing on $\mathcal{R}$ by
\begin{equation}
\langle A,B\rangle = A^{a\alpha}B^{b\beta}\mathcal{J}_{a\alpha,b\beta}
\end{equation}
For any complex field $X\in{\cal R}$ we define the gauge-covariant pseudoreal conjugate $\widetilde X={\cal J}X^*$ which, again, transforms in $\mathcal{R}$; for a spinor we similarly write $\widetilde\psi={\cal J}\psi^c$.

The matter content is:
\begin{table}[H]
\centering
\caption{Matter content of the CFT candidate.  All three fields transform in the same orthosymplectic bifundamental ${\cal R}$.}
\label{tab:OSpmatter}
\begin{tabular}{@{}lcccc@{}}
\toprule
field & statistics & $SO(2N_o)\times USp(2N_s)$ & $SU(3)_c$ & $(Q_{\text{EM}},Q_{B-L})$\\
\midrule
$Z^i$, $i=1,2,3$ & complex scalar & ${\cal R}$ & ${\bf3}$ & $(\frac16,\frac13)$\\
$Y$ & complex scalar & ${\cal R}$ & ${\bf1}$ & $(-\frac12,-1)$\\
$\psi$ & Dirac fermion & ${\cal R}$ & ${\bf1}$ & $(\frac12,0)$\\
\bottomrule
\end{tabular}
\end{table}
At the free point $\Delta_Z=\Delta_Y=1/2$ and $\Delta_\psi=1$. But, of course, we are actually interested in the ``Wilson-Fisher''-like fixed point that the theory flows to in the infrared. Preserving the global symmetry \eqref{eq:Fsym} requires a fine-tuning of the scalar potential, as discussed below.

The gauge-invariant fermionic operators
\begin{equation}
\boxed{
\begin{matrix}
 & &&\underline{(SU(3)_c,Q_{\text{EM}},Q_{B-L})}\\
 u^i&=\langle Z^{i},\psi\rangle,
 && \left({\bf3},\frac23,\frac13\right)\\
 d^i&=\langle Z^{i},\widetilde\psi\rangle,
 && \left({\bf3},-\frac13,\frac13\right)\\
 \nu&=\langle Y,\psi\rangle,
 && ({\bf1},0,-1),\\
 e&=\langle Y,\widetilde\psi\rangle,
 && ({\bf1},-1,-1).
\end{matrix}}
\label{eq:fourmesons}
\end{equation}
have exactly the right quantum numbers to correspond to the fermions of the first generation.
All four have the desired classical dimension
\begin{equation}
  \Delta_{\rm cl}=\frac32.
\end{equation}
corresponding to a massless bulk fermion.
These will receive corrections in the true IR fixed point of the CFT. However, as we will argue below, we expect the anomalous dimensions of the fermionic operators to be small.

With 3 complex fields we in principle have 3 independent $U(1)$ phase rotations as global symmetries.  One of those $U(1)$s will be explicitly broken by our scalar potential. In accord with the usual large-$N$ lore, we assume that the quartic part of the scalar potential (has been fine-tuned to) take the form of a sum of double-trace operators. The gauge-invariant scalar bilinears are listed in table \ref{tab:28scalars}.

\begin{table}[H]
\centering
\small
\caption{Gauge-invariant Scalar bilinears}
\label{tab:28scalars}
\begin{tabular}{lcccc}
\toprule
operator & $SU(3)_c$ & $Q_{\text{EM}}$ & $Q_{B-L}$ \\
\midrule
$M^i{}_j=\langle  Z^i, \tilde{Z}_j\rangle-\tfrac{1}{3}\delta^i{}_j\langle  Z^k, \tilde{Z}_k\rangle $ & ${\bf8}$ & $0$ & $0$\\
$N_1=\langle  Z^i, \tilde{Z}_i\rangle$ & ${\bf1}$ & $0$ & $0$ \\
$N_2=\langle Y,\tilde{Y}\rangle$ & ${\bf1}$ & $0$ & $0$ \\
$P_i=\frac12\epsilon_{ijk}\langle Z^j, Z^k\rangle$ & $\overline{\bf3}$ & $\frac13$ & $\frac23$ \\
$R^i=\langle Y, Z^i\rangle$ & ${\bf3}$ & $-\frac13$ & $-\frac23$ \\
$T_i=\langle Y, \tilde{Z}_i\rangle$ & $\overline{\bf3}$ & $-\frac23$ & $-\frac43$ \\
conjugates of $P,R,T$ & conjugate & opposite & opposite \\
\bottomrule
\end{tabular}
\end{table}

Out of these, we can form the $SU(3)\times U(1)^2$-invariant quartic potential
\begin{equation}
\begin{split}
V_4 &= a_1 M^i{}_j M^j{}_i+a_2 N_1^2+a_3 N_1N_2+a_4 N_2^2+a_5 P_i\overline{P}^i +a_6 R^i\overline{R}_i+a_7 T_i\overline{T}^i\\
&\qquad +bP_iR^i +b^* \overline{P}^i\overline{R}_i\\
\end{split}
\end{equation}
where the terms on the second line are responsible for breaking the ``extra'' $U(1)$ symmetry.

In accord with the usual large-$N$ dynamics, we expect the theory with $V_4$ turned on to flow to a fixed point. At the fixed point, the constituent scalars remain near $\Delta=1/2$, whereas scalar bilinears tend to be pushed up from $\Delta\simeq1$ to $\Delta\simeq2$. Turning on the Chern--Simons couplings (at least for small values of the 't Hooft couplings $\lambda_o=N_o/k_o$, $\lambda_s=N_s/k_s$), this means that the fermionic operators in \eqref{eq:fourmesons} remain near $\Delta=3/2$.

\subsection{Current contact terms}
\label{sec:newcontacts}
Three-dimensional  abelian conserved-current two-point functions can contain Chern--Simons contact terms,
\begin{equation}
  S_{\rm ct}^{(3d)}=\frac{\boldsymbol{k}^{IJ}_{\text{CFT}}}{4\pi}
  \int_{\partial\AdS_4} A_I\wedge \dd A_J+\cdots .
  \label{eq:3dcontactIJK}
\end{equation}
where we have introduced background gauge fields (sources) $A_I$.
The integer part of \(\boldsymbol{k}^{IJ}_{\rm CFT}\) is scheme-dependent, because it can be shifted by a local counterterm.

For the global form of the symmetry group
\begin{equation}
G= \frac{SU(3)\times U(1)_1}{\mathbb{Z}_3}\times U(1)_2
\end{equation}
the allowed counterterms have the form
\begin{equation}\label{CScounterterms}
\delta S = \frac{k}{4\pi} \int tr\bigl( A\wedge dA + \frac{2}{3} A^3 \bigr)+ \frac{\boldsymbol{k}^{IJ}_{\text{CT}}}{4\pi} \int A_I\wedge d A_J
\end{equation}
with
\begin{equation}\label{AllowedCScounterterms}
\boldsymbol{k} _{\text{CT}}=
\begin{pmatrix}
3k+9n_1& 3n_2\\
3n_2&k_2\\
\end{pmatrix}
\end{equation}
with $k,k_2,n_1,n_2\in\mathbb{Z}$.

Modulo the allowed counterterms, the value of $\boldsymbol{k}_{\text{CFT}}$ is invariant under RG flow and encodes the parity anomaly \cite{Closset:2012vp}. If nonzero, it must be reproduced by the bulk dual. In the present 3d CFT, $\boldsymbol{k}_{\text{CFT}}$ is easy to calculate from the UV (where the scalar potential is negligible). A fermion of charge $q^I$ under $U(1)_I$ contributes
\begin{equation}
\delta\boldsymbol{k}^{IJ} = -\tfrac{1}{2} q^I q^J
\end{equation}
In our model, the only fermion is $\psi$ in the bifundamental representation $\mathcal{R}=(\boldsymbol{2N}_o,\boldsymbol{2N}_s)$ of the gauge group. It has\footnote{All \emph{gauge-invariant} operators have integer charges under $U(1)_1\times U(1)_2$.} charge $(0,1/2)$ under $U(1)_1\times U(1)_2$. So it contributes
\begin{equation}
\begin{split}
\delta \boldsymbol{k}^{22} &= -\dim(\mathcal{R})\cdot \tfrac{1}{2} \left ( q^{I=2} \right )^2\\
&= - \frac{N_oN_s}{2}
\end{split}
\end{equation}
If $N_oN_s$ is even, this vanishes mod $\mathbb{Z}$. If not, we would need bulk $\theta$-terms to reproduce the parity anomaly.

\subsection{The full single-cycle spectrum}
\label{sec:fullspectrum}

The economy of Table~\ref{tab:OSpmatter} does not imply a sparse single-particle spectrum.  The theory is a matrix large-$N$ theory.  Writing a bifundamental as $X^{a\alpha}$, a connected gauge invariant is an even-length cycle with alternating $SO$ and $USp$ contractions, for example
\begin{equation}
  {\cal O}[X_1,\ldots,X_{2\ell}]
  =\Tr_{SO}\!\left[
    (X_1\Omega X_2^T)(X_3\Omega X_4^T)\cdots
    (X_{2\ell-1}\Omega X_{2\ell}^T)
  \right].
  \label{eq:unorientedcycle}
\end{equation}
The orthosymplectic invariants identify a cycle with its orientation reversal, so these single-trace operators are naturally unoriented necklaces.  Products of disconnected cycles are multi-trace operators and correspond to multiparticle states.

Already the scalar bilinears in table \ref{tab:28scalars}  consist of 28 real components: $8+1+1+6+6+6=28$.  We expect these to have conformal dimensions $\sim$few. And we have a whole tower of other single-trace operators, as we just saw. This is what we expect in a large-N CFT, and it invariably signals that, in the bulk,  the AdS${}_4$ is accompanied by a compact manifold $M$ whose size is not \emph{parametrically} small in units of the AdS radius. All we really know about $M$ is that it has no continuous isometries (which would, upon Kaluza-Klein reduction, give rise to additional gauge symmetries of the bulk theory).

This was not specific to the particular 3d CFT we proposed; it is a generic feature of any large-$N$ CFT. As such, we can say that, by stabilizing the Higgs potential at some sub-Planckian AdS minimum, we have made the great desert bloom.

To understand the phenomenological impact of these extra neutral and colored scalars what is important is not directly their mass in the high-VEV vacuum but their mass in our vacuum, which is not just determined by their bare masses but also by their coupling to both the Higgs and the $B-L$ breaking scalar $\phi$. For any one of the extra scalars their mass is given by
\begin{equation}
M^2(|H|^2,|\phi|^2) = m_0^2 + \kappa_1 |H|^2 + \kappa_2 |\phi|^2 + \ldots
\end{equation}
where the terms in the ellipses come from non-renormalizable terms in the Lagrangian and so are Planck suppressed. $\kappa_1$ and $\kappa_2$ are in principle determined from the CFT, but require a computation not just of anomalous dimensions but of actual 4-pt functions.

As long as $\kappa_1$ and $\kappa_2$ are of order 1 the mass in our vacuum is entirely driven by $\kappa_1$ and furthermore $\kappa_1$ has to be negative\footnote{Properly, as in \S\ref{particleMasses}, we should use the running coupling constants $\kappa_{1,2}$, evaluated at the high/low VEV. For this crude estimate, that will not make a difference.}.  The difference between the mass squared in our vacuum and the high scale vacuum is 
\begin{equation}
\Delta M^2 = \tfrac{\kappa_1}{2} (v_1^2 - v_2^2) + \tfrac{\kappa_2}{2} v_{\phi}^2 \approx - \tfrac{\kappa_1}{2} v_2^2
\end{equation}
where in the last step we used $v_2 \gg v_\phi \gg v_1$. Multiplying by $L^2$, this says that the mass${}^2$ in our vacuum
\begin{equation}
M^2L^2 \approx\Delta(\Delta+3) -\kappa_1\frac{108c_6}{|\lambda_*|^2}
\end{equation}
which would be tachyonic if $\kappa_1$ were positive. The metastability of our vacuum requires $\kappa_1<0$. If the CFT 4-pt function were to predict a positive $\kappa_1$, that CFT would be ruled out immediately. On the other hand for $\kappa_1$ negative, the extra scalars will be very heavy in our vacuum and so phenomenologically harmless. Their contribution to the running of the Higgs coupling will be negligible below the AdS scale.

Without a detailed proposal for the geometry of the extra dimensions to match against not much can be learned from the generic structure of the ``KK" operators, that is the various operator families we can construct by inserting neutral scalar or fermion bi-linears into the cycle of \eqref{eq:unorientedcycle}. But let us briefly discuss a few particularly simple neutral insertions. First consider the color-singlet bilinears \begin{equation} {\cal S}_Z=(Z_j^\dagger Z^j), \qquad {\cal S}_Y=Y^\dagger Y, \label{eq:neutralinsertions} \end{equation} where in the expressions below these bilinears are inserted inside a single connected gauge-index cycle rather than multiplied as separate gauge-invariant traces. For example, the up-type sector contains the short operators \begin{equation} \begin{aligned} u_{(1)}^i&= \langle Z^{i}, \psi \rangle , \\ u_{(2)}^i&= \bigl[Z^i\psi\,{\cal S}_Z\bigr], 
\\ u_{(3)}^i&= \bigl[Z^i\psi\,{\cal S}_Y \bigr],
\end{aligned} \label{eq:upfamilycandidates} \end{equation} 
with classical dimensions \begin{equation} \Delta_{\rm cl} = \frac32,\quad \frac52,\quad \frac52 . \end{equation} 
The square brackets indicate connected cycles as in \eqref{eq:unorientedcycle}.
These operators seem to be natural candidates to describe the 3 visible generations of standard model fermions. There will be more generations at larger dimensions, presumably too heavy to have been observed in nature. Mixing between the two $\Delta=5/2$ candidates would have to push the third generation up to its large anomalous dimension.

Last but not least, beyond the particles observed  in the standard model, the particular variant of the great desert scenario our CFT is supposed to be dual to also included the $B-L$ breaking scalar $\phi$ with $B-L$ charge two and color and E\&M neutral. Our CFT does have a natural operator with these properties, 
\begin{equation}
O_{\phi} =  \bigl[Y \psi Y \psi \bigr]^\dagger 
\end{equation}
of classical dimension 3, presumably closer to 4 after taking into account that it's quadratic in the scalars. For this to match with \eqref{eq:phimass} we either need to have $c_6$ and $\lambda_3$ that are small-ish, or the operator $O_{\phi}$ has to acquire a large anomalous dimension.
Overall, our CFT candidate has all the low lying operators required to match the spectrum in table \ref{tab:mLdimensions} together with a large tower of extra states necessitating an extra dimension of size comparable to the AdS scale.

\section{Discussion and Future Directions}
\label{sec:discussion}

The main point we tried to make in this work is that the great desert would be far from a disappointment: with the additional assumption of a stable minimum it predicts that quantum gravity in our world has an exact UV completion in terms of a 3d CFT that gives unique predictions for all particle masses and couplings. Clearly finding this CFT dual to the great desert would be a huge theoretical advance.

Much can be learned about the properties of the putative dual from what we know about the Standard Model in our vacuum simply by running couplings up to high energy.  At minimum the CFT has the global symmetry $SU(3)_c\times U(1)_{\rm EM}$, while in the neutrino branch pursued here it also has $U(1)_{B-L}$.  Parity is broken.  The CFT should contain fermionic operators with dimensions very near $3/2$ to account for the light matter fields, together with a large stress-tensor normalization.  The candidate in Sec.~\ref{sec:cftcandidate} realizes these quantum numbers with four short fermion--scalar mesons and, because it is a matrix large-$N$ theory, also contains an infinite set of connected single-trace operators.  It would be extremely interesting to see to what extent the conformal bootstrap constrains CFTs with these symmetries and this low-lying spectrum.

The proliferating operator spectrum suggest that the bulk has at least one large extra dimension. We believe this feature to be quite generic. Typical large $N$ gauge theories will have towers of states that require accounting for. To make progress one needs to find ways to pin down the full operator spectrum in the CFT, which is of course challenging due to strong coupling effects.

Last but not least, it would be very interesting to understand whether the orthosymplectic CFT and the associated higher dimensional bulk admit a natural string-theory embedding.  Supersymmetric $SO\times USp$ Chern--Simons theories are known to arise as orientifold relatives of M2-brane constructions~\cite{Hosomichi:2008jb} suggesting that at least one of the extra dimensions should be an interval bounded by two orientifold planes.

\section*{Acknowledgments}
We are grateful to Kristan Jensen and Can Kilic for helpful discussions. This work was supported in part by DOE grant DE-SC0022021 and by a grant from the Simons Foundation (Grant 651678, AK). JD would like to thank the Simons Summer Workshop at the Simons Center for Geometry and Physics, Stony Brook University, where some of this work was performed.
Some of the calculations in this work have been performed with help of Claude Code and GPTpro.

\bibliographystyle{utphys}
\bibliography{orthobiblio}
\end{document}